\documentclass[twocolumn,aps,prl,superscriptaddress,showpacs]{revtex4-1}
\usepackage[utf8]{inputenc}
\usepackage{color}
\usepackage{amsmath}
\usepackage{amssymb}
\usepackage{graphicx}
\usepackage{esint}
\usepackage[unicode=true,
 bookmarks=false,
 breaklinks=true,pdfborder={0 0 0},backref=false,colorlinks=true]
 {hyperref}
\usepackage{xcolor}
\makeatletter

\usepackage{epstopdf}\usepackage{subfigure}
\usepackage{float}
\usepackage{xcolor}
\usepackage{amsfonts}
\usepackage{bm}
\usepackage{subfigure}

\makeatother

\begin{document}

\title{A diagnosis scheme for topological degeneracies at high-symmetry momenta}

\author{Tiantian Zhang}
\email{zhang.t.ac@m.titech.ac.jp}
\affiliation{Institute
of Physics, Chinese Academy of Sciences/Beijing National Laboratory for Condensed Matter Physics, Beijing 100190, China}
\affiliation{Department of Physics, Tokyo Institute of Technology, Ookayama, Meguro-ku, Tokyo 152-8551, Japan}
\affiliation{Tokodai Institute for Element Strategy, Tokyo Institute of Technology, Nagatsuta, Midori-ku, Yokohama, Kanagawa 226-8503, Japan}
\author{Ling Lu}
\affiliation{Institute
of Physics, Chinese Academy of Sciences/Beijing National Laboratory for Condensed Matter Physics, Beijing 100190, China}
\author{Shuichi Murakami}
\affiliation{Department of Physics, Tokyo Institute of Technology, Ookayama, Meguro-ku, Tokyo 152-8551, Japan}
\author{Zhong Fang}
\affiliation{Institute
of Physics, Chinese Academy of Sciences/Beijing National Laboratory for Condensed Matter Physics, Beijing 100190, China}
\affiliation{Collaborative Innovation Center of Quantum Matter, Beijing, 100084,
China}
\author{Hongming Weng}
\email{hmweng@iphy.ac.cn}
\affiliation{Institute
of Physics, Chinese Academy of Sciences/Beijing National Laboratory for Condensed Matter Physics, Beijing 100190, China}
\affiliation{Collaborative Innovation Center of Quantum Matter, Beijing, 100084,
China}
\author{Chen Fang}
\email{cfang@iphy.ac.cn}
\affiliation{Institute
of Physics, Chinese Academy of Sciences/Beijing National Laboratory for Condensed Matter Physics, Beijing 100190, China}
\affiliation{CAS Center for Excellence in Topological Quantum Computation, Beijing 100190, China}

\begin{abstract}

Theories of symmetry-based indicators and topological quantum chemistry, while powerful in diagnosing gapped topological materials, cannot be directly applied to diagnosing band degeneracies at high-symmetry momenta due to the violation of the compatibility conditions. Here we design a recursive protocol that utilizes indicators of maximal subgroups to infer the topological nature of band degeneracies at high-symmetry lines. For demonstration, the method is used to predict the existence of iso-energy Weyl points and a node-line cage, respectively, in the phonon bands of In$_2$Te and ZrSiO.

\end{abstract}

\maketitle

\paragraph*{Introduction}

\begin{center}
\begin{figure}
\includegraphics[scale=2]{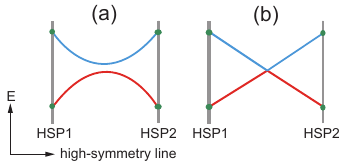}\caption{Two different kinds of band structure along a high-symmetry line. The red line represents the valence band, and the blue line represents the conduction band. HSP1 and HSP2 represent two different high-symmetry point in the Brillouin zone. The band structure in (a) satisfies the compatibility condition, while (b) does not.\label{fig:CC}}
\end{figure}
\end{center}

Theories of symmetry-based topological indicators\cite{po2017symmetry} and topological quantum chemistry\cite{bradlyn2017TQC} are useful in diagnosing topological gapped materials such as topological insulators\cite{song2018quantitative,song2018diagnosis,Top_insulator_2,Top_band} and topological crystalline insulators\cite{zhang2019catalogue,vergniory2019complete,tang2019comprehensive}. The application of the theory presupposes that along high-symmetry lines in the Brillouin zone (BZ), the conduction and the valence bands do not cross each other, or, put in a formal way, the valence bands satisfy the compatibility conditions, {as shown in Fig.~\ref{fig:CC}(a)}. When these conditions are violated for any pair of high-symmetry points, one only knows that the system is gapless, $i. e.$, having band crossing between the conduction and the valence bands, {as shown in Fig.~\ref{fig:CC}(b)}. Aside from the existence, the method does not yield further information on the band topology associated with these degeneracies.

The importance of such information, however, is highlighted in the research of topological semimetals, such as Weyl semimetals\cite{weng2015weyl,fang2012multi,DoubleWeyl_1,DoubleWeyl_2,Weyl_newfermions}, Dirac semimetals\cite{Top_CDAS,Top_NA3BI,Dirac_1,Dirac_2}, and nodal-line semimetals\cite{xu2015two,fang2016topological}. The nontrivial topology of band degeneracies in some of these materials leads to unique surface states having ``Fermi arcs''\cite{Weyl_Taas,Weyl_experiments,Weyl_exp,DoubleWeyl_3} as well as an anomalous bulk transport phenomenon known as the ``quantum anomaly''\cite{son2013chiral,huang2015observation,xiong2015evidence,zyuzin2012topological,nielsen1983adler}. Topologically nontrivial band degeneracies appear not only in electronic bands but are also predicted and observed in the bands of bosons such as photons\cite{lu2014topological,lu2016topological,lu2015experimental,TOP_photonic4}, phonons\cite{zhang2018double,miao2018observation,Top_phon_mirco1,Top_phon_mirco2,Top_phon_mirco3,Top_phon_micro4,Top_phon_micro5,Weyl_phon_mecha,Top_acoustic,Weyl_acoustic4,Top_huber1,Top_HUBER2,Top_phonon6,TP_XUYONG,xia2019symmetry}, and magnons\cite{li2017dirac,yao2018topological}.

\begin{center}
\begin{figure}
\includegraphics[scale=1]{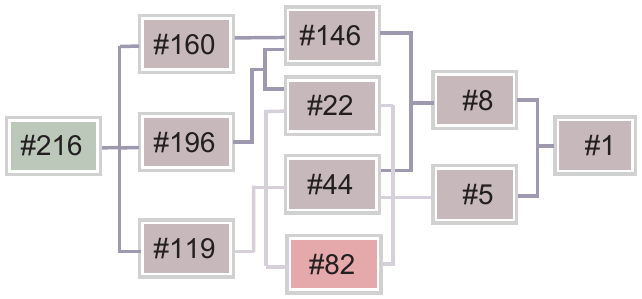}\caption{Tree-search process for \#216. \#82 is the maximum subgroup which has a nontrivial symmetry-based indicator and satisfies compatibility condition.\label{fig:Diag_method}}
\end{figure}
\end{center}

In this Letter, we develop a routine for topological-diagnosing band degeneracies located along high-symmetry lines in BZ, especially for orthogonal Hamiltonians\cite{schnyder2008classification,chiu2016classification,ryu2010topological}. Given a band structure that violates the compatibility conditions of some space group $G$, we first identify, through a ``tree-search'' process, the {maximum} subgroup $H\subset{G}$, such that the compatibility conditions of $H$ are satisfied, and then compute the symmetry-based indicators with respect to $H$. The values of these subgroup indicators then reveal partial information on topological invariants of the band degeneracies protected by $G$, which has not been discussed in detail before\cite{song2018diagnosis,vergniory2019complete,tang2019comprehensive}. For example, the lowest six phonon bands of In$_2$Te violate the compatibility conditions of \#216. Fig.~\ref{fig:Diag_method} shows all subgroups of space group \#216, from which we identify \#82 as the {maximum} subgroup that has nontrivial indicator group ($\mathbb{Z}_2$) and satisfies the compatibility condition at the same time. We then compute the $\mathbb{Z}_2$-indicator for \#82 and find $z_2=1$, which by Ref.[\onlinecite{song2018diagnosis}] ensures that the band degeneracies between the sixth and the seventh band are Weyl points of equal energy. We also apply the method to the phonon bands of ZrSiO and show the presence of a ``nodal cage'' in its band structure.

\begin{figure*}
\includegraphics[scale=0.92]{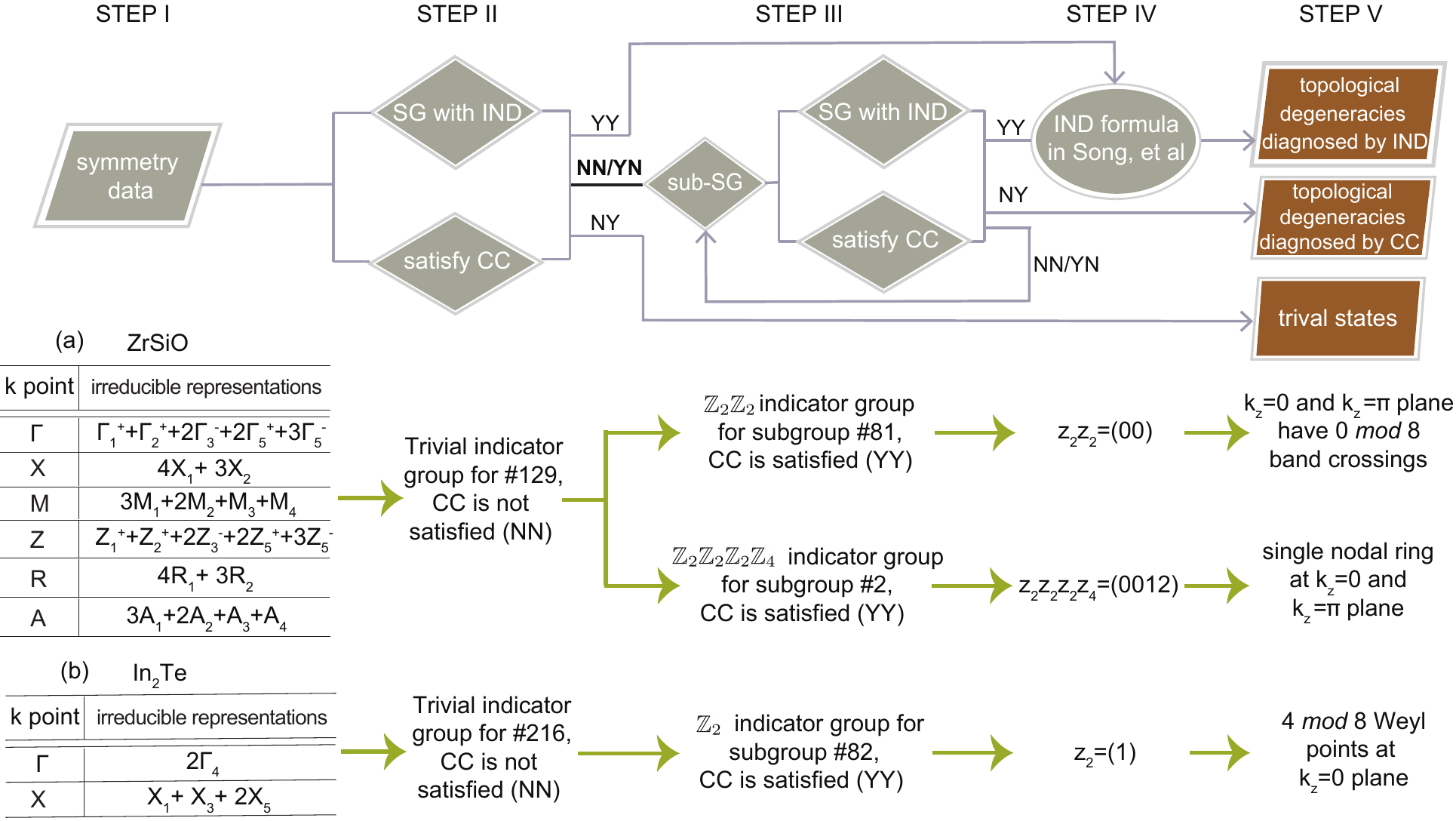}\caption{Calculation steps for diagnosing topological degeneracies at high-symmetry momenta. SG=space group, IND=indicator, CC=compatibility condition, N=No, Y=Yes. First of all, symmetry data should be obtained by first-principle calculations, and fed into the next step. Secondly, check whether the space group $G$ has a nontrivial indicator and the symmetry data satisfies compatibility conditions. If both the answers are yes, we can use the symmetry data to calculate the indicator directly and get the information of topological degeneracies at generic momenta for AI class systems. If the answers are ``NY'', the material is a topologically trivial one. Otherwise, topological degeneracies will exist at high-symmetry momenta, such as high-symmetry points, high-symmetry lines, and high-symmetry planes. To get the complete information of the topological band crossings, we should find a maximum subgroup $H$ in step III, which has a nontrivial symmetry-based indicator group and satisfies compatibility condition at the same time. After calculating the symmetry-based indicator for $H$ in step IV, we can get the information of topological degeneracies for space group $G$. 
However, if we can not find a subgroup $H$ in step III, then the topological degeneracies only can be diagnosed by compatibility condition.
 (a) In$_{2}$Te (b) ZrSiO are two examples for demonstrating our diagnosis scheme for AI class systems.  \label{fig:flowchart} }
\end{figure*}

\paragraph*{Flowchart for the recursive algorithm}{}%
The diagnosing process for topological degeneracies at high-symmetry momenta is summarized in Fig.~\ref{fig:flowchart}, which consists of five steps.
In step I, symmetry data, $i.e.$, irreducible representation, at a given list of high-symmetry momenta should be calculated by first-principle calculations. 
In step II, analyze if two conditions, $i.e.$, the nontrivial-symmetry-based-indicator-group condition (INDC) and compatibility condition~(CC) are satisfied by the space group and the symmetry data, respectively.
If the space group $G$ has a nontrivial symmetry-based indicator group, and the symmetry data satisfies CC, then we can calculate the symmetry-based indicator directly by Ref.[\onlinecite{song2018diagnosis}] to get the information of topological band degeneracies at generic momenta in the BZ.
If space group $G$ does not have a nontrivial symmetry-based indicator group, but the symmetry data satisfies CC, then the system is in a ``trivial'' state.
(Here ``trivial'' means the symmetry data is that of an atomic insulator.)
In other cases, CC is not satisfied, $i.e.$, band degeneracies will exist at high-symmetry momenta, so we should use a new method to diagnose the information for band crossings in the BZ.
In Step III, we find all the maximal subgroups of $G$, and choose one $H$ from this set, and test if the INDC and CC are satisfied: if the answer is YY, we proceed to step IV; if the answer is YN or NN, we replace $G$ by $H$ and repeat step III; if the answer is NY, we replace $G$ by another one of its maximal subgroup $H'\neq{H}$ and repeat step III until all the maximal subgroups have been exhausted.
In step IV, we should use the symmetry-based indicator formula of $H$ to diagnose the topological degeneracy information of space group G, no matter the indicator is zero or not.
After step IV, replace $G$ by another one of its subgroups $H'\neq{H}$ unless all maximal subgroups are exhausted.
In the last step, the indicator of subgroup $H$ can tell us the information of topological degeneracies at high-symmetry momenta, such as types, configuration, positions, and topological charge for the topological degeneracies in the BZ.
If the process never reaches step IV but stops when all the maximal subgroups of $G$ are considered, we conclude that the nodes cannot be further diagnosed by eigenvalues at all (non-diagnosable).
After the process, each band nodes protected by $G$ are to be found either as non-diagnosable, or as having a list of subgroups with corresponding indicators, which carry topological information on these nodes.

\begin{center}
\begin{figure*}
\begin{widetext}
\includegraphics[scale=0.9]{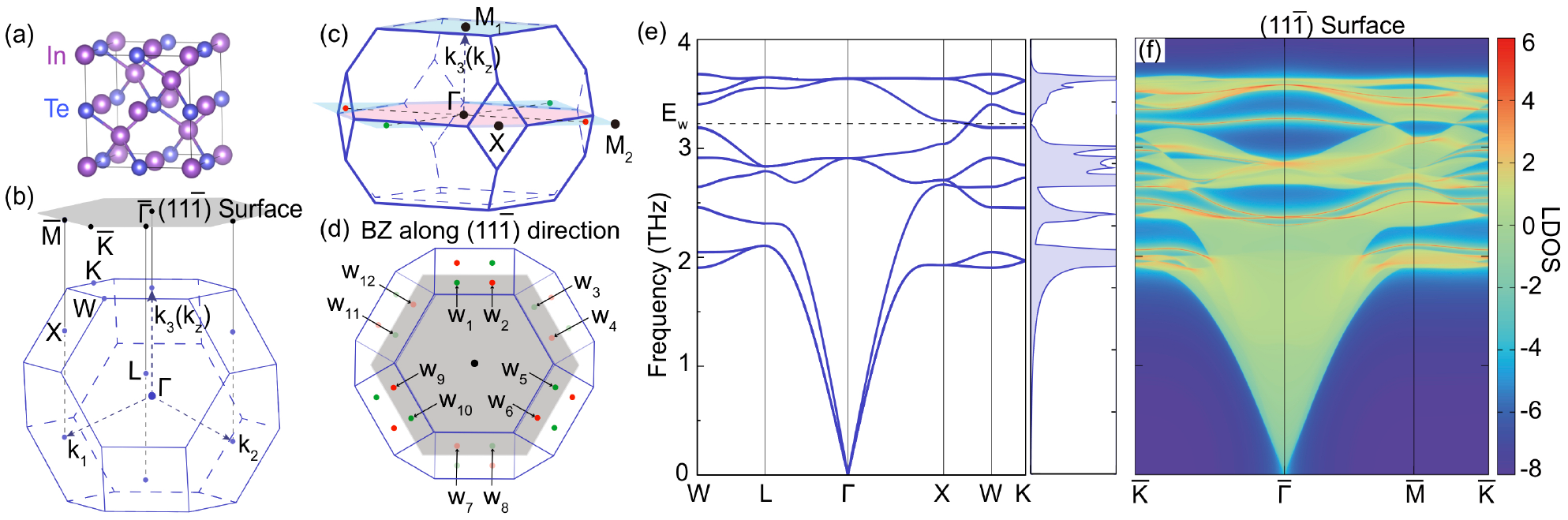}
\end{widetext}
\caption{(a) Crystal structure for In$_{2}$Te, (b) Brillouin zone and surface Brillouin zone along $(11\bar{1})$ direction for \#216.
(c) Brillouin zone for \#82. (d) Distribution for Weyl points in the Brillouin zone, where green dots represent Weyl points with Chern number of $-1$ and red dots represent Weyl points with Chern number of +1.
(e) Phonon spectra for In$_{2}$Te and (f) is the surface state along $(11\bar{1})$  direction. \label{fig:In2Te}}
\end{figure*}
\end{center}

Figure.~\ref{fig:flowchart} also list two materials for the demonstration of our recursive algorithm in phononic systems.
 (a) is an example of In$_{2}$Te, which has a noncentrosymmetric structure and isoenergy Weyl points in the phonon spectra. (b) is an example of ZrSiO, which has a centrosymmetric structure and nodal cage band degeneracies in the phonon spectrum. Both of the cases break CC in step II.

\paragraph*{Diagnosing process for In$_{2}$Te}
In$_{2}$Te belongs to a $\mathcal{P}$-broken space group $F\bar{4}3m$ (\#216)\cite{In2B_crystal}, as shown in Fig.\ref{fig:In2Te}(a). The band crossing at around 3.22THz indicates that CC is broken along $X-W$ direction when the number of bands is 6, as shown in Fig.~\ref{fig:In2Te}(e). In the following, we will get the complete information for the band crossing at around 3.22THz by using the diagnosing method shown in Fig.~\ref{fig:flowchart}.
 
After obtaining symmetry data for \#216 in step I\cite{CAL_DFPT,CAL_VASP}, we notice that CC for \#216 is not satisfied along $X-W$ direction, which is not a surprise because of the band crossing in phonon spectra. However, there is not a nontrivial indicator for space group \#216 in step II. Therefore in step III, we map each irreducible representation from \#216 to \#82~($I\bar{4}$), which is the maximum subgroup having a nontrivial indicator $\mathbb{Z}_{2}$ and satisfying CC. 
In step IV, we calculate the topological invariant of subgroup \#82 and get a nonzero indicator $z_{2}=1$, which indicates that there will be 4 $mod$ 8 Weyl points on the $k_{z}=0$ plane (which is also $k_{3}=0$ plane). In the following, we will provide an intuitive perspective to understand how to get the complete information for the topological degeneracies at high-symmetry momenta for \#216 from the indicator of subgroup \#82.

Figure.~\ref{fig:In2Te}(c) shows the BZ for space group \#82, which has a similar shape with the one for \#216 shown in Fig.~\ref{fig:In2Te}(b). Since there are two pieces of plane for $k_3=0$ plane in Fig.~\ref{fig:In2Te}(b), we can rebuild the BZ by dividing the blue quadrilateral plane into four pieces and fill them into the pink plane separately to get a new quadrilateral $k_{1}=0$ plane. In this case, high-symmetry point $M_{1}$ in the old BZ is $M_{2}$ in the new BZ. 
Since there are 4 $mod$ 8 Weyl points on the $k_{1}=0$ plane, shown by red and green dots in Fig.~\ref{fig:In2Te}(c), we find that there will be 12 $mod$ 24 Weyl points in the BZ after considering all the symmetry operators of \#216, especially $C_{3}$ symmetry around (111) direction. A detailed calculation confirms that there are 12 robust Weyl points related to each other by symmetries at around 3.22THz, confined in in $k_{1}=0$, $k_{2}=0$, and $k_{3}=0$ planes. 

\paragraph*{Surface states for the Weyl points in In$_{2}$Te}
Fig.~\ref{fig:In2Te}(d) show the positions of 12 Weyl points in the BZ, where green dots represent left-hand Weyl points with Chern number of $-1$ and red dots represent right-hand Weyl points with Chern number of $+1$.
 Among those six pairs of Weyl points, $W_{1}$ and $W_{2}$ have opposite chirality, which are related by mirror symmetry $m_{011}$. The Weyl pairs $W_{1,2}$, $W_{5,6}$, and $W_{9,10}$ are related to each other by $C_{3}$ rotation symmetry along (111) direction, and those three pairs of Weyl points are related with the remaining three pairs of Weyl points by $\mathcal{T}$. 
Local density of states (LDOS) on the $(11\bar{1})$ surface is shown in Fig.~\ref{fig:In2Te}(f), plotted along the surface mometum lines $\bar{K}-\bar{\Gamma}-\bar{M}-\bar{K}$. The surface arcs are clearly seen at around 3.22THz.

Here, we note that there are another 12 Weyl nodes at around 2.58THz, with the same indicator $z_{2}=1$. Surface states and surface arcs for the 12 Weyl nodes at around 2.58THz are discussed in the supplementary materials.
 
\begin{center}
\begin{figure}
\includegraphics[scale=0.75]{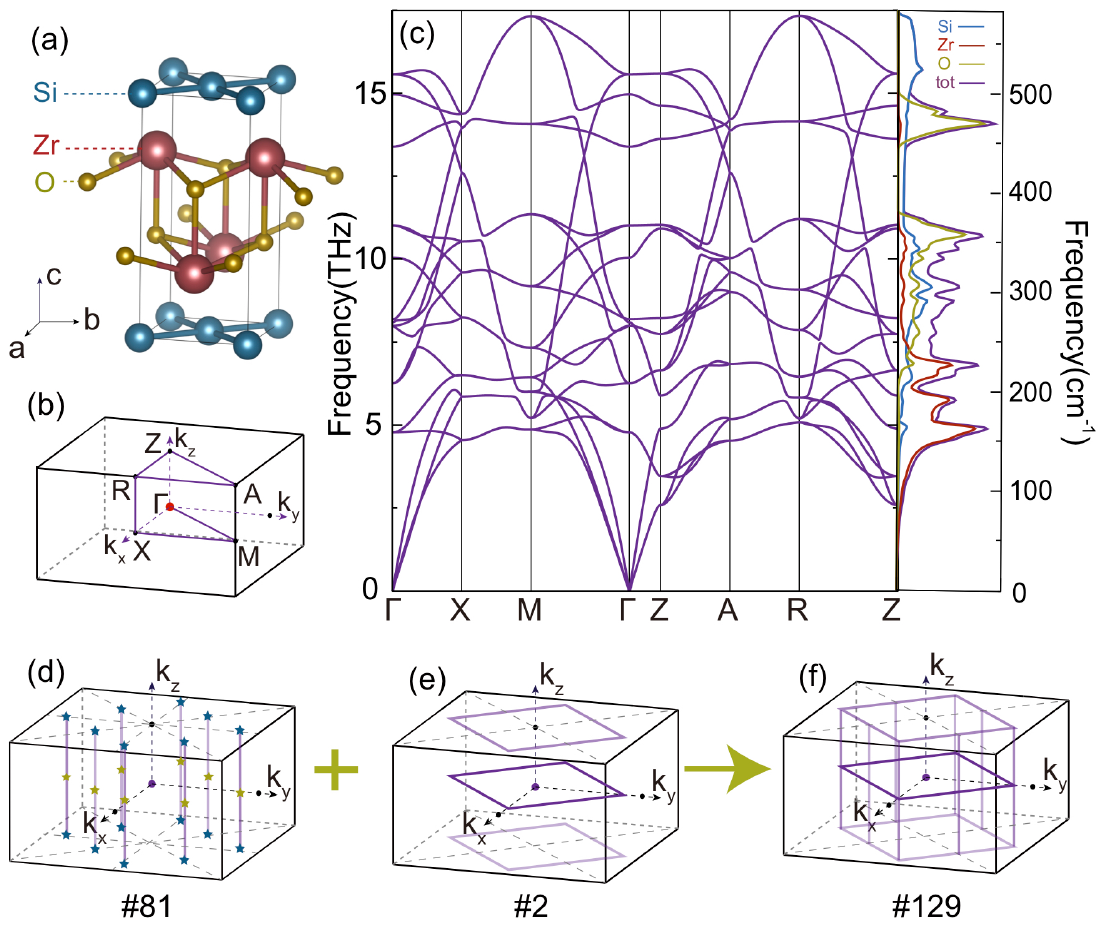}\caption{(a) Crystal structure, (b) Brillouin zone, and (c) phonon spectra for ZrSiO. (d-f) are the configurations for node-line degeneracies in Brillouin zone. All the nodal lines carry a quantized $\pi$ Berry phase. \label{fig:FIG_2} }
\end{figure}
\end{center}

 \paragraph*{Diagnosis process for ZrSiO}
ZrSiO has a centrosymmetric structure with space group $P4/nmm$ (\#129)\cite{onken1964silicid}, as shown in Fig.~\ref{fig:FIG_2}(a).
ZrSiO is a nodal line semimetal for electronic structures when spin-orbit coupling is ignored\cite{xu2015two} and features the same topological properties for phonon spectra.
 Figure.~\ref{fig:FIG_2}(c) shows the phonon bands for ZrSiO, which have several band crossings at around 13THz along several high-symmetry lines. We will discuss what they are, and whether they are topological protected in the following.

After obtaining the symmetry data by density-function perturbation theory in step I, we find out that they do not satisfy the CC along $\Gamma-X$, $\Gamma-M$, $Z-A$, and $Z-R$ directions when the occupied bands is 14 in step II. Violation of CC indicates that a band degeneracy will exist between the 14th and 15th band at those four high-symmetry lines, which also means we can use the recursive algorithm to figure out the complete topological information for the degeneracies. 
In step III, we find that the maximum subgroup having a nontrivial symmetry-based indicator for \#129 is \#85. However, CC is not satisfied for \#85. Thus, we need to proceed the recursive process to check the subgroups with lower symmetries. After iterating several times, we get two maximum subgroups having a nontrivial indicator and satisfying CC, which are \#81 and \#2. \#81 and \#2 belong to different ``tree branches'', which means that neither of them is a subgroup of the other one.

The indicator group for \#81 is $\mathbb{Z}_{2}\times\mathbb{Z}_{2}$, and the corresponding topological invariants for phonons in ZrSiO are $(00)$. Even though the topological invariants are zero, they can still tell us that there will be 0 $mod$ 8 band crossings at both $k_{z}=0$ plane and $k_{z}=\pi$ plane. As shown in \ref{fig:FIG_2}(d), band crossings are marked by pentagrams.
 Since $(\mathcal{PT})^{2}=1$ in ZrSiO, all the band crossings must belong to nodal lines/rings.
Therefore a possible configuration for the band degeneracies in ZrSiO diagnosed by \#81 is 0 $mod$ 8 nodal lines crossing both $k_{z}=0$ plane and $k_{z}=\pi$ plane along (001) direction, as shown in Fig.~\ref{fig:FIG_2}(d).

The indicator group for \#2 is $\mathbb{Z}_{2}\times \mathbb{Z}_{2}\times \mathbb{Z}_{2}\times \mathbb{Z}_{4}$, and the corresponding topological invariants are (0012) for phonons in ZrSiO.
 Those nonzero indicators tell us a new configuration for the topological degeneracies in BZ, and we offer two different perspectives to understand it. 
(i) Because indicator groups are Abelian groups, indicators satisfy the sum rule. Thus (0012) can be written as (0001)+(0011), which indicates a single nodal ring around $\Gamma$ and $Z$ point by Ref.[\onlinecite{song2018diagnosis}], respectively. $M_{z}$ symmetry in ZrSiO will restrict those two nodal rings in the $k_{z}=0$ and $k_{z}=\pi$ plane, which corresponds to the configuration shown in  Fig.~\ref{fig:FIG_2}(e). (ii) $z_{2}=1$ and $z_{4}=2$ indicate that any curved surface in the BZ passing $\Gamma$, $X$, and $M$ point (or $R$, $A$, and $Z$ point) will be crossed by nodal lines 2 $mod$ 4 times. Therefore one possible configuration is one nodal ring lying on the $k_{z}=0$ and $k_{z}=\pi$ plane respectively, which also matches Fig.~\ref{fig:FIG_2}(e) well. 

In the last step, we can get the complete configuration for topological degeneracies in BZ by combining the indicators of subgroup \#81 and \#2.  Further calculations show that there are 8 nodal lines and 2 nodal rings in the BZ, $i.e.$, node-cage degeneracies shown in Fig.~\ref{fig:FIG_2}(f), which perfectly matches the diagnosis.
Indicators from those two subgroups also tell us that each nodal line/ring in ZrSiO carries a $\pi$ Berry phase, and zero monopole charge. Surface states for nodal lines/rings are in a drum-head shape, but it is not easy to observe them experimentally for being easily covered by the bulk dispersions. 

\paragraph*{Conclusion}
We designed a recursive algorithm for diagnosing the complete information of topological degeneracies locating at high-symmetry momenta by using the indicators of maximum subgroups. 
This recursive algorithm can be used in any systems with $\mathcal{T}^{2}=1$ and compensates for the shortcomings of the previous method, which can only be used in systems satisfying compatibility conditions, $i.e.$, no band degeneracies at high-symmetry momenta. In two examples for phonon bands, $i.e.$, In$_{2}$Te and ZrSiO, we diagnosed all the topological degeneracies in the BZ successfully, which verifies the effectiveness of the diagnosing methd.

%


\bibliographystyle{unsrt}
\bibliography{reference}
 \newpage{}
\newpage{}

\end{document}